\documentclass[column,secnumarabic,amssymb,amsmath,nofootinbib,floatfix,
nobibnotes,aps,
superscriptaddress,
prc
]{revtex4}

\usepackage[dvipsnames]{xcolor}
\usepackage{graphicx}%
\usepackage{amsmath}
\usepackage{slashed}
\usepackage{subcaption}
\usepackage{bm}

\begin{document}
\title{Relativistic meson-exchange currents in semi-inclusive lepton scattering}

\author{V. Belocchi} 
\affiliation{
	Dipartimento di Fisica, Universit\`a di Torino, via P. Giuria 1, 10125 Turin, Italy}
\affiliation{
	INFN, Sezione di Torino, via P. Giuria 1,10125, Turin, Italy
}
\author{M.B. Barbaro} 
\affiliation{
	Dipartimento di Fisica, Universit\`a di Torino, via P. Giuria 1, 10125 Turin, Italy}
\affiliation{
	INFN, Sezione di Torino, via P. Giuria 1,10125, Turin, Italy
}
\author{A. De Pace} 
\affiliation{
	INFN, Sezione di Torino, via P. Giuria 1,10125, Turin, Italy
}
\author{M. Martini}
\affiliation{IPSA-DRII,  63 boulevard de Brandebourg, 94200 Ivry-sur-Seine, France}
\affiliation{Sorbonne Universit\'e, Universit\'e Paris Diderot, CNRS/IN2P3, Laboratoire
de Physique Nucl\'eaire et de Hautes Energies (LPNHE), Paris, France}

\begin{abstract}
We assess the impact of two-particle--two-hole excitations on the semi-inclusive electron scattering process $(e,e'p)$ using a fully relativistic nuclear model calculation that precisely incorporates antisymmetrization. The calculation encompasses all contributions involving the exchange of a single pion and the excitation of a $\Delta$ resonance. Our results are compared with $(e,e'p)$ data on carbon at kinematics where two-nucleon emission dominates. This work represents an essential step towards the microscopic computation of the two-particle--two-hole contribution to semi-inclusive neutrino reactions, crucial in the analysis of neutrino oscillation experiments.
\end{abstract}

\maketitle

\section{Introduction}
Several measurements of neutrino-nucleus semi-inclusive  cross sections have been recently performed \cite{T2K:2018rnz,MINERvA:2018hba,MINERvA:2019ope,MicroBooNE:2020fxd,MicroBooNE:2020akw,MicroBooNE:2023tzj}, with the aim of reducing the systematic error associated to nuclear effects in the analysis of long baseline neutrino oscillation experiments. In particular, the control of nuclear effects is crucial to guarantee the success of  future neutrino experiments (HyperK 
\cite{Hyper-Kamiokande:2018ofw} and DUNE \cite{DUNE:2015lol}) in the search of CP violation in the leptonic sector.
 
Semi-inclusive measurements, corresponding to the simultaneous detection of the outgoing lepton and one or more hadrons in the final state, are indeed more sensitive to the details of nuclear modeling compared to inclusive cross sections, which only involve the detection of the outgoing lepton.
While most theoretical studies have predominantly focused on charge-current (CC) inclusive $(\nu_l,l)$ reactions, recent efforts have expanded to investigate the semi-inclusive reaction $(\nu_l,lN)$ in the quasi-elastic (QE) channel, corresponding to the scattering of the probe with a single nucleon~\cite{Moreno:2014kia,Franco-Patino:2022tvv,Franco-Patino:2023msk}. 

It is now well-established that two-particle--two-hole (2p2h) excitations provide a substantial contribution to the inclusive $(\nu_l,l)$ cross section without pions in the final state. Several calculations of this contribution are available~\cite{Martini:2009uj,Nieves:2011pp,RuizSimo:2016rtu,VanCuyck:2017wfn,Rocco:2018mwt,Lovato:2020kba}. Despite presenting some differences \cite{Katori:2016yel}, 
all of them agree with the inclusive data within the error bars. The situation differs for the semi-inclusive neutrino reaction $(\nu_l,lN)$, for which a complete microscopic calculation of the 2p2h contribution is still absent in the literature\footnote{So far the only semi-inclusive $(\nu_l,lN)$ published results including 2p2h excitation are the ones of the Ghent group \cite{VanCuyck:2016fab,VanCuyck:2017wfn}, which, however, do not include the $\Delta$ contributions to meson exchange currents. Semi-inclusive $(\nu_l,lNN)$ cross sections at fixed neutrino energies have also been recently calculated \cite{Martinez-Consentino:2023hcx}.}, although it is strongly needed for the correct interpretation of neutrino data. Currently the 2p2h component of the cross section is simulated in event generators on the basis of {\it inclusive} calculations \cite{Nieves:2011pp,RuizSimo:2016rtu}, following a procedure which necessarily relies on some strong approximations. Inclusive processes involve a summation over all accessible intermediate nuclear states, consequently predicting cross sections only as functions of the outgoing lepton kinematics. In principle, such approach cannot be employed to predict semi-inclusive or exclusive cross sections, where a specific hadronic final state is detected. However, 
given the lack of theoretical calculations of these contributions, the  strategy taken so far has been to "extract" exclusive predictions from inclusive results, forcibly using assumptions \cite{Dolan:2019bxf, Hayato:2021heg} whose reliability is difficult to control. 
In general, the resulting 2p2h contribution represents a significant component of the semi-inclusive cross section, as shown in Refs. \cite{Franco-Patino:2022tvv, Franco-Patino:2023msk} for neutrinos and in Ref. \cite{CLAS:2021neh} for electrons. Consequently, not only is the implementation of the 2p2h itself questionable, but it also has implications for the conclusions drawn regarding nuclear models used to describe the one-body nuclear response.
For instance, the $(\nu_\mu, \mu p)$ cross section, recently measured by T2K \cite{T2K:2018rnz}, MINERvA \cite{MINERvA:2019ope}, and MicroBooNE \cite{MicroBooNE:2020fxd}, has been shown to be highly sensitive to the model used to describe the final state interactions (FSI) between the ejected proton and the residual nucleus. Different FSI prescriptions can yield significantly different results in general \cite{Franco-Patino:2022tvv, Franco-Patino:2023msk}, and the comparison of theoretical predictions with data is strongly influenced by the contribution of 2p2h events to the experimental signal.

The only correct approach to implement a model for 2p2h in a Monte Carlo generator for the semi-inclusive reaction is through a microscopic calculation of these contributions.
As a first step towards this goal, in this work we consider the semi-inclusive electron scattering reaction $(e,e'p)$. As for inclusive scattering, validation with semi-inclusive electron scattering data is an essential benchmark for any nuclear model intended for use in neutrino scattering studies. We hence generalize to the semi-inclusive $(e,e'p)$ process the previous inclusive fully relativistic antisymmetrized calculation of 2p2h electromagnetic responses performed in Ref.~\cite{DePace:2003spn}, already extended to the weak sector \cite{RuizSimo:2016rtu} and implemented in the GENIE Monte Carlo generator~\cite{Dolan:2019bxf}. 
The results of Ref. \cite{DePace:2003spn}, coupled to the "SuSAv2" model~\cite{Gonzalez-Jimenez:2014eqa} for the one-body response, provide a very good description of electron scattering inclusive data across a broad kinematic range~\cite{Megias:2016lke} as well as of all the available inclusive neutrino cross sections~\cite{Megias:2016fjk,Megias:2017cuh,Megias:2018ujz,Amaro:2019zos},  with the exception of the lower momentum and energy transfer regime where collective nuclear excitations dominate.

Several high quality $(e,e'p)$ datasets have been collected by past experiments \cite{Lourie:1986zz,Baghaei:1989cc,Kester:1995wkm,Kester:1996mp,JeffersonLabHallA:2004pgx} 
 and compared to theoretical calculations \cite{Ryckebusch:1994pi,Ryckebusch:1997gn,Giusti:1994lxi}, 
in kinematic conditions where multi-nucleon emission is expected to give the largest contribution to the cross section - the so-called dip region between the QE and $\Delta$ production peaks. The comparison with these data is particularly useful for application to neutrino scattering studies: indeed the broad neutrino energy distribution typical of oscillation experiments, in contrast with the very precisely known beam energy in electron experiments, does not allow for a clear separation of the different reaction channels, which contribute to the same experimental signal.

\section{Formalism}
\label{sec:Formalism}
Let us consider the $(e,e'N)$ process
\begin{eqnarray}
e+A \to e'+N+X 
\label{eq:eepN}
\end{eqnarray}
where an electron of energy $\varepsilon$ scatters off a nucleus $A$ at rest in the laboratory frame, transferring to it an energy $\omega=\varepsilon-\varepsilon'$ and a momentum ${\bf q}={\bf k}-{\bf k'}$~\footnote{We work in the extreme relativistic limit (ERL) $\varepsilon\gg m_e$, implying $k=\varepsilon$ and $k'=\varepsilon'$, and we neglect the nuclear recoil.}. In the final state, the electron has energy $\varepsilon'$ and scattering angle $\theta_e$, and a nucleon of energy $E'_1$, momentum $p_1$ and solid angle $\Omega'_{1}$ is knocked out, leaving a residual system $X$. 
The 6th-differential cross section for this process can be written in terms of four nuclear response functions as
\begin{widetext}    
\begin{equation}
\frac{d^6\sigma}{d\omega d\Omega'_{e} dE'_1 d\Omega'_{1}} = 
\frac{p_1 E'_1}{(2\pi)^3} \sigma_M \left[ v_L R_L^{(N)}+v_T R_T^{(N)}+v_{LT} R_{LT}^{(N)} \cos\phi_1'+v_{TT} R_{TT}^{(N)} \cos(2\phi_1') \right]
\label{eq:d6sig}
\end{equation}
\end{widetext}
where 
$\sigma_M = \frac{\alpha^2\cos^2(\theta_e/2)}{4\varepsilon^2\sin^4(\theta_e/2)}
$
    is the Mott cross section, $\phi_1'$ is the nucleon's azimuthal angle and
\begin{eqnarray}
v_L &=& \frac{Q^4}{q^4} \,,
\ \ 
v_T = \frac{Q^2}{2q^2}+\tan^2\frac{\theta_e}{2} \,,
\\
v_{LT}  &=& \frac{Q^2}{q^2} + \sqrt{\frac{Q^2} {q^2}+\tan^2\frac{\theta_e}{2}}\,,
\ \ 
v_{TT}   = \frac{Q^2}{2q^2}
\end{eqnarray}
are kinematic leptonic factors, with $Q^2=q^2-\omega^2$. 

The response functions $R^K\equiv R^K(q,\omega,{\bf p_1})$ are specific components of the semi-inclusive hadronic tensor $W^{(N)}_{\mu\nu}$, which encodes the information on the nuclear structure and currents. Specifically:
\begin{eqnarray}
R_L^{(N)} &=& W_{00}^{(N)}
\label{eq:rl}\\
R_T^{(N)} &=& W_{xx}^{(N)}+W_{yy}^{(N)}
\label{eq:rt}\\
R_{LT}^{(N)} \cos\phi_1' &=& -W_{0x}^{(N)}-W_{x0}^{(N)}
\label{eq:rlt}\\
R_{TT}^{(N)}  \cos(2\phi_1') &=& W_{xx}^{(N)}-W_{yy}^{(N)} \,.
\label{eq:rtt}
\end{eqnarray}

The nuclear tensor for the $(e,e'N)$ process \eqref{eq:eepN} is given by
\begin{equation}
    W_{\mu\nu}^{(N)} = \sum_X \langle A|\hat J_{\mu}^{\dagger}|N,X\rangle \langle N,X| \hat J_\nu | A\rangle
    \,\delta\left(E_N+E_X-E_0-\omega\right) \,,
    \label{eq:Wmunu}
\end{equation}
where $|A\rangle$ denotes the nuclear ground state having energy $E_0$,  $|N,X\rangle$ is the hadronic final state of energy $E_N+E_X$ and a sum is performed over the unobserved states $X$. 

The nuclear current in Eq.~\eqref{eq:Wmunu}, $\hat J^\mu =  \hat J^\mu_{1b} +\hat J^\mu_{2b}$, is the sum of one- and two-body currents, neglecting higher order contributions.
As a consequence, the hadronic tensor can be expressed as the sum of various contributions, corresponding to the excitation of different final states, one-particle-one-hole (1p1h) and two-particle-two-hole (2p2h):
\begin{equation}
    W_{\mu\nu}^{(N)} = W_{\mu\nu}^{(N) \rm 1p1h}+W_{\mu\nu}^{(N) \rm 2p2h}\,.
\end{equation}

In this work we focus on the 2p2h hadronic tensor $W_{\mu\nu}^{(N) \rm 2p2h}$, corresponding to the ejection of two nucleons from the nucleus, in order to test the validity of our meson exchange currents (MEC) model. For a general kinematics it could be arduous to separate, in the experimental data, 2p2h contributions from the Quasi-Elastic (QE) process, which occurs at lower transferred energy, or from pion production.
However, for the  kinematics analyzed in this work, the QE responses are very small and pion production can affect the cross-section only at missing energies higher than the pion threshold, as we show in the results.

We analyze the semi-inclusive reaction in which one nucleon (a proton) is detected in the final state while the second nucleon (a proton or a neutron) is unobserved. The inclusive reaction, in which none of the two outgoing nucleons are detected, was extensively studied in Ref.~\cite{DePace:2003spn} in the Relativistic Fermi Gas (RFG) framework and we adopt here the same formalism and model used in that reference.

Within the RFG, the semi-inclusive 2p2h hadronic tensor reads (from now on we shall omit for brevity the label 2p2h)

\begin{equation}
 W_{\mu\nu}^{(N)} := \frac{d\,W_{\mu\nu}}{d{\bf p_1}}=\frac{1}{4}  \frac{V}{(2\pi)^9}  \sum_{\substack{s_1,s_2,s'_2 \\t_1,t_2,t'_2}} \int
\,d{\bf h_1} \,d{\bf h_2} \,d{\bf p_2} \, \Theta_{\rm PB}\, \delta^4(p_1+p_2-h_1-h_2-\tilde q)\,w_{\mu\nu}
\label{eq:dW}
\end{equation} 
$$\Theta_{\rm PB} := \theta(k_F-|{\bf h_2}|)\,\theta(k_F-|{\bf h_1}|)\,\theta(|{\bf p_1}|-k_F)\,\theta(|{\bf p_2}|-k_F) \qquad \tilde q := (\tilde \omega,\mathbf{q}), \qquad \tilde \omega:= \omega - E_s^{2p2h}$$
where $V=3\pi^2 A/2k_F^3$ is the nuclear volume, $h_i$ and $p_i$ are the initial and final four-momenta of the two nucleons with on-shell relativistic energies $E_i$ and $E'_i$,$E_s^{2p2h}$ is a parameter that accounts for the nucleus energy absorption in the 2p2h process, $k_F$ is the Fermi momentum, the factor $1/4$ is needed to avoid double counting for indistinguishable particles and holes in the final state and 
\begin{equation}
     w_{\mu\nu} = \frac{1}{16 E_1 E_2 E'_1 E'_2} \langle F|\hat J_\mu^{2b\dagger}|2p2h\rangle \langle 2p2h | \hat J_\nu^{2b} | F \rangle 
    \label{eq:Wtilde}  
\end{equation}
is the tensor containing reduced matrix elements.
The kets $|F\rangle$ and $|2p2h\rangle$ represent the ground state and the two-particle-two-hole excitation, respectively.
After some manipulation 
the reduced matrix element present in Eq.~\eqref{eq:Wtilde} can be written as:
\begin{align}
     \langle 2p2h | \hat J^\mu_{2b} | F \rangle
    &=j^\mu(h_1,h_2,p_1,p_2)-j^\mu(h_1,h_2,p_2,p_1) \,.
\end{align}
Hence there are two different terms in the explicit expression of the matrix element, that we can call \emph{normal ordered} (NO) and \emph{inverted ordered} (IO).
The product of two matrix elements appearing in Eq.~\eqref{eq:Wtilde} leads to four terms: thanks to 
symmetry properties it is possible to group these terms in two different contributions, called \emph{direct} and \emph{exchange}:
\begin{eqnarray}
    w^{\mu\nu} = \frac{1}{16 E_1 E_2 E'_1 E'_2}\big[2\underbrace{j^{\mu \dagger}_{\rm NO}\,j^{\nu}_{\rm NO}}_{\rm direct} -  \underbrace{j^{\mu \dagger}_{\rm NO}\,j^{\nu}_{\rm IO}}_{\rm exchange} - \underbrace{j^{\mu \dagger}_{\rm IO}\,j^{\nu}_{\rm NO}}_{\rm exchange}\big] \,.
\end{eqnarray}

In the following subsections we provide explicit expressions for the two-body currents used in this work. 

\subsection{Meson Exchange Current}

\begin{figure*}
\includegraphics[scale=0.3]{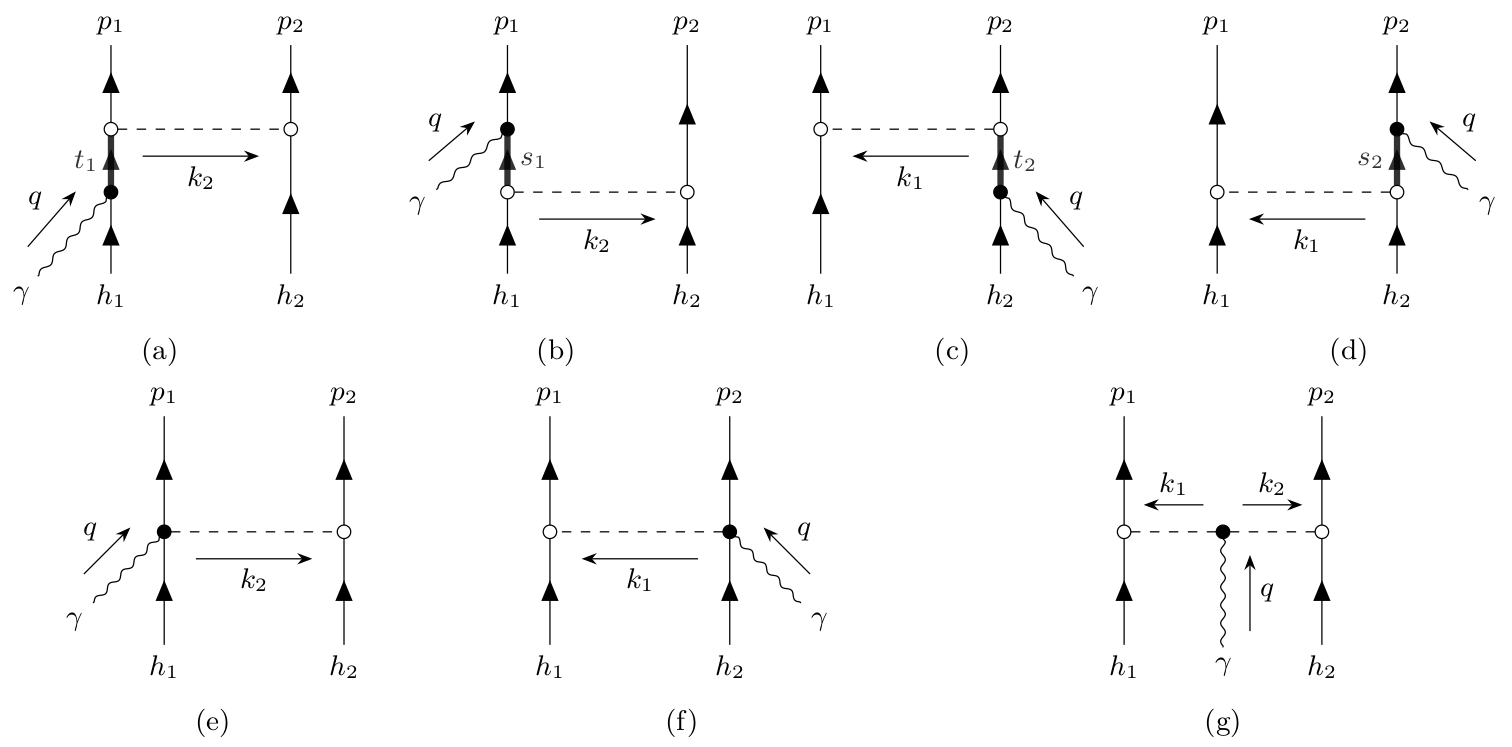}
\caption{First row: $\Delta$-MEC diagrams, forward (a,c) and backward (b,d). The $\Delta$ resonance is represented by the thick line. Second row: pure pionic diagrams, seagull (e,f) and pion-in-flight (g). Note that $h_i$ stand for the initial nucleon state, $p_i$ for the outgoing nucleon. The photon vertex is the black solid dot, while internal nuclear vertexes are the white ones.}
\label{fig:MEC}
 \end{figure*}

Our computation involves fully relativistic two-body currents $J_\mu^{2b}$ corresponding to the coupling of the virtual photon to a pair of nucleons exchanging a pion. It is obtained starting from the Non-Linear $\sigma$-model Lagrangian investigated in \cite{Hernandez:2007qq}, where the $\Delta$ resonance is included \emph{ad hoc}. The corresponding diagrams in free space, shown in Fig.~\ref{fig:MEC}, are usually denoted as pion-in-flight, $j^\mu_\pi$, seagull (or contact), $j^\mu_{\rm sea}$, and $\Delta$-MEC, $j^\mu_\Delta$. The latter involves the excitation of an intermediate $\Delta$ resonance and gives the dominant contribution to the 2p2h cross section. 
 Therefore these different contributions appear separately in the total MEC operator: 
 $$j^\mu_{\rm MEC}=j^\mu_\pi +j^\mu_{\rm sea}+j^\mu_\Delta\,.$$
Note that $j^\mu$ is an operator in the isospin space. 
 
The pure pionic part of the Electro-Magnetic (EM) MEC is characterized by the exchange of a charged $\pi$ between the two nucleon currents. This appears explicitly in the isospin operator associated to $j^\mu$:
$$j^\mu_\pi=I_{V_3}J^\mu_\pi \qquad \qquad j^\mu_{\rm sea}=I_{V_3}J^\mu_{\rm sea}$$
\begin{align}
J^\mu_\pi &= \frac{f^2_{\pi NN}}{m_\pi^2}F_1^V(q) F_{\pi NN}(k_1)F_{\pi NN}(k_2)(k_1^\mu-k_2^\mu)\bar{u}(p_1)\gamma_5 \slashed{k_1} u(h_1)\bar{u}(p_2)\gamma_5 \slashed k_2 u(h_2)\big(\Delta_\pi(k_1)\Delta_\pi(k_2)\big)^*\
\label{eq:pif}\\
J^\mu_{\rm sea}&= \frac{f^2_{\pi NN}}{m_\pi^2}F_1^V(q) F^2_{\pi NN}(k_1) \bar{u}(p_1)\gamma_5 \slashed{k_1} u(h_1)\Delta_\pi(k_1)\bar{u}(p_2)\gamma_5 \gamma^\mu u(h_2) - (1\leftrightarrow 2)
\label{eq:seag}
\end{align}
$u(h_i)$ are the nucleon isospinor, $F_1^V(q)$ form factor accounts for the photon EM coupling, $F_{\pi NN}$ is the hadronic form factor for the $\pi NN$ coupling with the $\pi$ off-shell and $\Delta_\pi$ the pion propagator. The $*$ in the pion-in-flight currents stands for the gauge preserving procedure applied, showed in \cite{Dekker:1994yc}.
For details, see Appendix \ref{appendix pion}.

The $\Delta$-MEC include those diagrams in which a nucleon state is excited into a $\Delta$.
There are two different kind of excitations, called \emph{forward} and \emph{backward}, depending on the relative position of the virtual $\Delta$ respect to the photon coupling vertex, appearing in two different kind of diagrams in Fig. \ref{fig:MEC}. 
$$j^\mu_\Delta=j^\mu_{\Delta_{\rm F}}+j^\mu_{\Delta_{\rm B}}$$
\begin{equation}
    j^\mu_{\Delta_{\rm F}}= I_{\Delta_{\rm F}}\frac{f^*f_{\pi NN}}{\sqrt{6} m_\pi^2} F_{\pi N\Delta}(k_1)F_{\pi NN}(k_1) \bar{u}(p_1)\gamma_5 \slashed{k_1} u(h_1)\Delta_\pi(k_1) k_1^\alpha  \bar{u}(p_2) G_{\alpha \beta}(t_2)\Gamma^{\beta \mu}(h_2,q)u(h_2)+ (1\leftrightarrow 2)
    \label{eq:deltaF}
\end{equation}
\begin{equation}
    j^\mu_{\Delta_{\rm B}}= I_{\Delta_{\rm B}}\frac{f^*f_{\pi NN}}{\sqrt{6}m_\pi^2} F_{\pi N\Delta}(k_1)F_{\pi NN}(k_1) \bar{u}(p_1)\gamma_5 \slashed{k_1} u(h_1)\Delta_\pi(k_1) k_1^\alpha  \bar{u}(p_2) \widetilde{\Gamma}^{\mu \beta}(p_2,q)G_{\beta \alpha }(s_2)u(h_2)+ (1\leftrightarrow 2)
        \label{eq:deltaB}
\end{equation}
with $t_i=h_i+q$, $s_i=p_i-q$. $f^*$ and $F_{\pi N \Delta}$ are respectively the coupling constant and the form factor describing the $\pi N \Delta$ vertex, while $\Gamma^{\mu \alpha}$ and $\tilde{\Gamma}^{\alpha \mu}$ are the form factors for the $\gamma N \Delta$ interaction. $\Delta$ appears as a virtual resonance only, and $G_{\alpha \beta}$ is its Rarita-Schwinger propagator.
$I_\Delta$ is the isospin operator, which is different in the two case. 
\\It is very useful to divide the operator $J_\Delta$ in term of isospin operators ($I_{\Delta}$ is a combination of them), writing the full MEC current as 
\begin{equation}
    j^\mu_{\rm MEC}=I_{V_3}(J^\mu_\pi+J^\mu_{\rm sea}+J^\mu_{\Delta_3})+2\tau_3^{(1)} J^\mu_{\Delta_1}+2\tau_3^{(2)} J^\mu_{\Delta_2}\,.
\end{equation}
For details and isospin separated resonance currents see Appendix \ref{appendix delta}.
\section{Results}
\label{sec:Results}

\begin{figure}
    \includegraphics[scale=0.35]{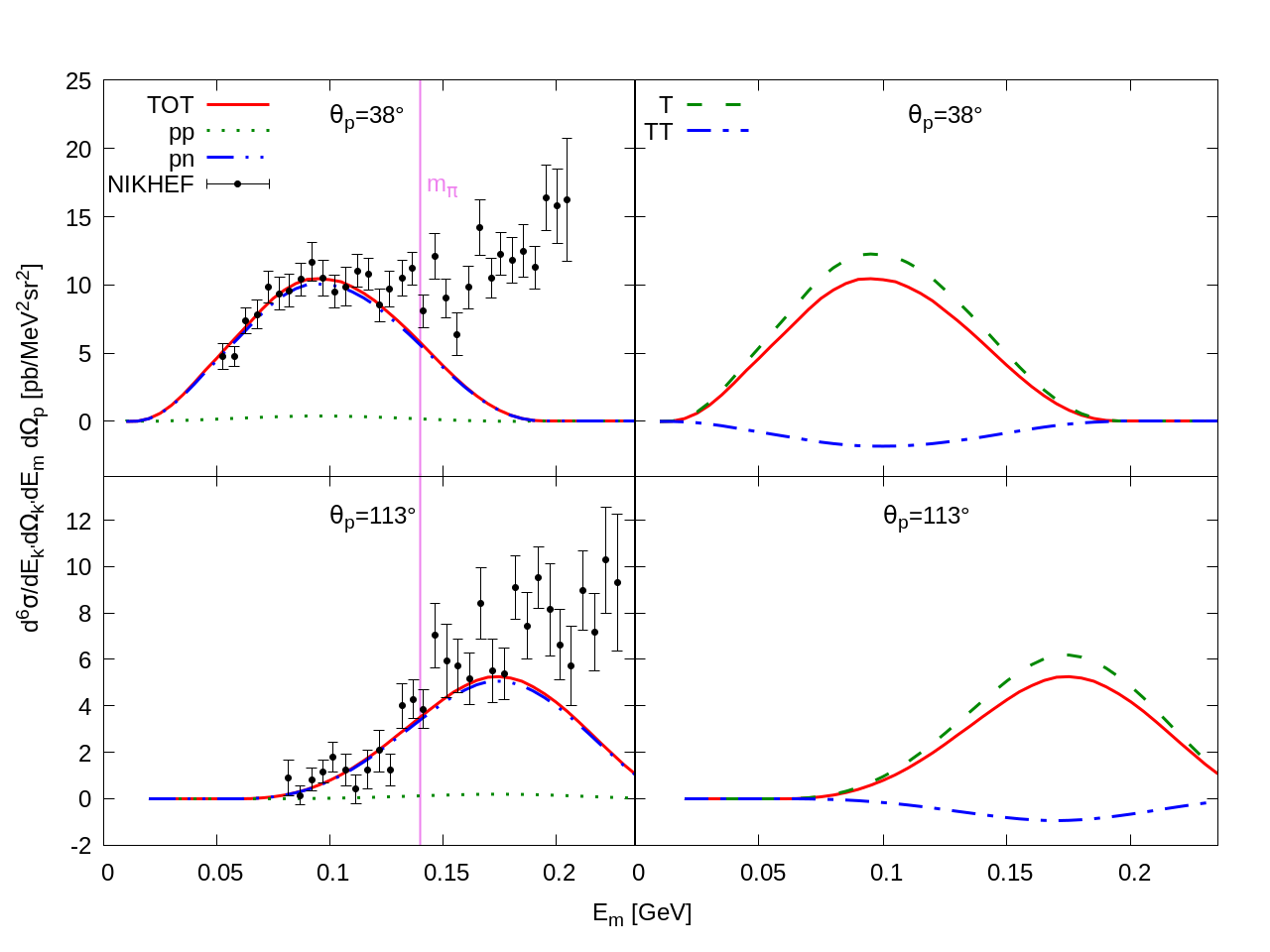}
    \caption{The 2p2h contribution to the $^{12}C(e,e'p)$ cross section is displayed versus the missing energy for electron energy $\epsilon$=478 MeV, energy transfer $\omega$=263 MeV, momentum transfer $q$=303 MeV/c and proton scattering angle $\theta_p$=38$^\circ$, 113$^\circ$ degrees. 
    The separate contributions of $pp$ and $pn$ emission to total cross section is shown in the left panels, and the $\pi$ production threshold is the vertical line in violet.
    The T and TT contributions are shown separately in the right panels. Data from Ref.~\cite{Ryckebusch:1994pi}.}
    \label{fig:NIKHEF1}
\end{figure}
\begin{figure}
 \begin{center}
    \includegraphics[scale=0.35]{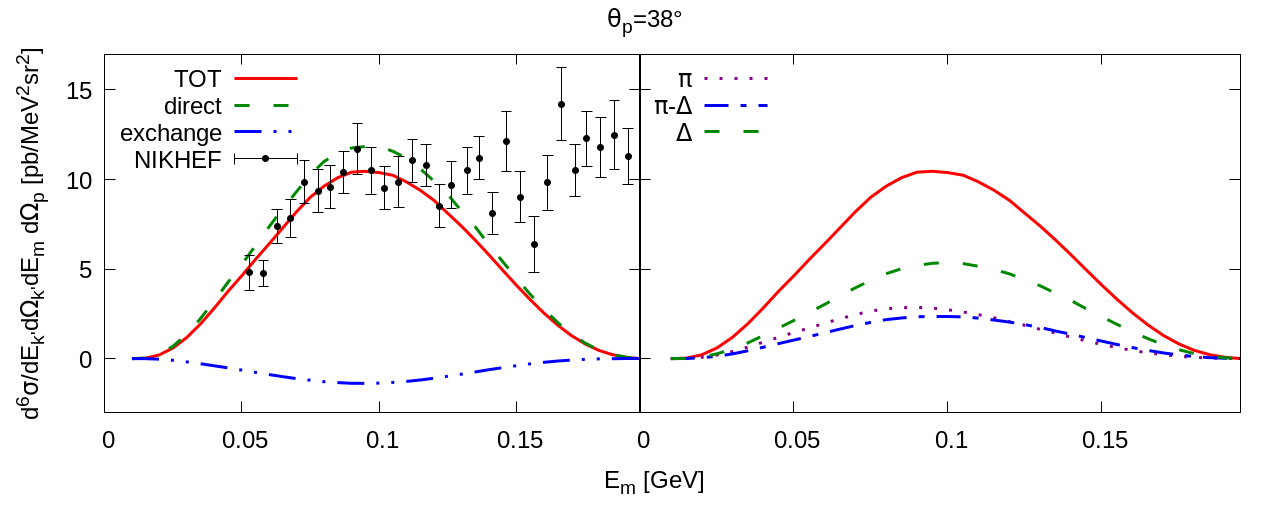} 
 \end{center}  
    \caption{The 2p2h contribution to the $^{12}C(e,e'p)$ cross section is displayed versus the missing energy for the same kinematics of Fig.~\ref{fig:NIKHEF1} and proton scattering angle $\theta_p$=38$^\circ$. Left panel: the separated contributions of the direct and exchange diagrams. Right panel: separated contributions of the $\pi$, $\Delta$ and $\pi-\Delta$ interference. Data from Ref.~\cite{Ryckebusch:1994pi}. }
\label{fig:NIKHEF2}
\end{figure}
\begin{figure*}
    \centering
  \includegraphics[scale=0.35]{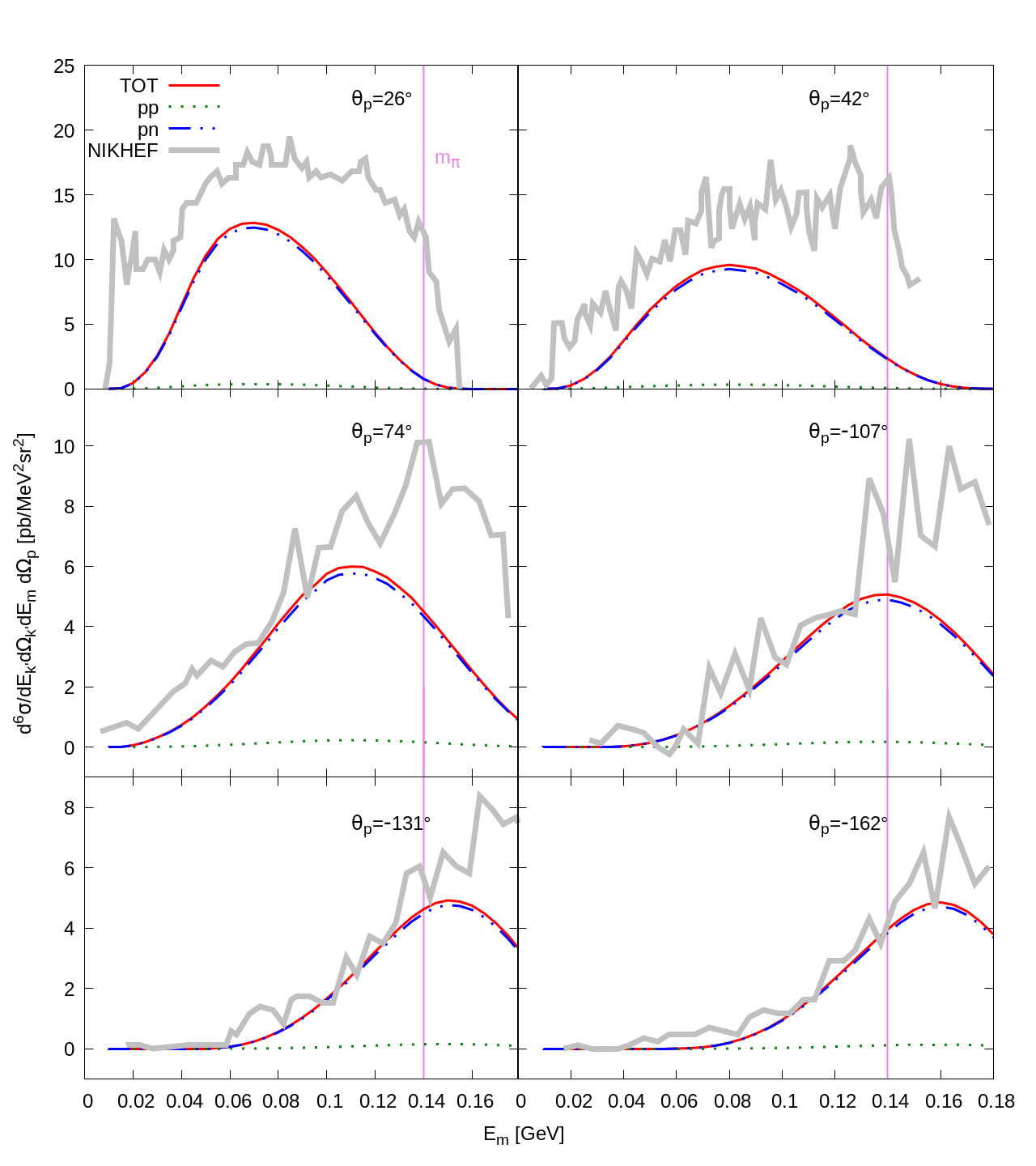 }
    \caption{The 2p2h  contribution to the $^{12}C(e,e'p)$ cross section is displayed versus the missing energy for electron energy $\varepsilon$=475 MeV, energy transfer $\omega$=212 MeV, momentum transfer $q$=270 MeV/c and several proton scattering angles $\theta_p$=26°, 42°, 74°, 107°, 131° and 162°. Data are taken from Ref.~\cite{Kester:1995wkm}. Experimental uncertainties are not shown. The pion threshold  is displayed as a vertical line.} 
    \label{fig:NIKHEF3}
\end{figure*}

We now present the results of the calculation described in the previous section and we compare our predictions with $(e,e'p)$ data from NIKHEF~\cite{Kester:1995wkm, Kester:1996mp, Ryckebusch:1994pi, Ryckebusch:1997gn} and MIT-Bates~\cite{Baghaei:1989cc}.
\\
The evaluation of the hadronic tensor involves a huge amount of terms, emerging from the MEC operator. In Eq.~\eqref{eq:dW} we fixed the detected proton momentum, $p_1$, evaluating the hadronic tensor for each possible configuration of the two holes and the remaining particle, yielding to a nine-dimension integration, but its possible to cut down the computational effort through some manipulation.
Thus the integral has been partially performed exploiting the energy and momentum conservation $\delta$ functions, removing the integration over $p_2$ and $\theta_2$ the hole polar angle, reducing it to a five-fold integration over two hole momenta, azimuthal angles and one polar angle. It has been evaluated using the CUBA library \cite{Hahn_2005}, that provides a tool for multidimensional Monte Carlo integration. We computed the contributions of 4 direct pionic diagrams, 12 (6 direct + 6 exchange) $\Delta$ resonance ones and the same amount for the $\pi-\Delta$ interference. 

Our starting point is the RFG model, which contains two parameters, the Fermi momentum $k_F$ and the energy shift $E_s$, fitted to the width and position of the $(e,e')$ quasi-elastic peak. For the $^{12}C$ nucleus we use $k_F$=228 MeV/c and $E_s$=20 MeV, according to Ref.~\cite{Maieron:2001it}. The energy shift phenomenologically accounts for the nucleon binding energy and for final state interaction effects. In the present calculation we replace $\omega\to\tilde \omega$ in the energy-conserving $\delta$ function of Eq.~\eqref{eq:dW}, using, unless differently stated, $E_s^{2p2h}= 2E_s$ = 40 MeV. Note that this modification affect the nuclear state description only, but not the EM interaction, being the EM nucleon form factor $F_1^V$ evaluated using $\omega$. The Hoeler parametrization for $F_1^V$ has been adopted~\cite{Hohler:1976ax}. Although more modern parametrizations of the form factors exist \cite{PhysRevC.70.068202,Bradford_2006}, this choice is motivated by the comparison  with the results of Ref.~\cite{DePace:2003spn}, that we follow also for the pionic and Delta form factors. It should be noticed that the analyzed data correspond to very low $Q^2$ values, below $0.1$ GeV$^2$ and we have checked that the results are almost insensitive to the form factor parametrizations.

The 6th differential cross sections, evaluated using Eq.~\eqref{eq:d6sig}, are plotted as functions of the missing energy
\begin{equation}
    E_m = \omega - T_p \,,
\end{equation}
$T_p$ being the kinetic energy of the knocked-out proton, for fixed values of the electron energy $\varepsilon$, energy transfer $\omega$, momentum transfer $q$ and angle $\theta_p$ between the proton direction and the momentum transfer ${\bf q}$. As in Ref.~\cite{Ryckebusch:1994pi}, we neglect the contribution of the L and LT responses, shown in Eqs.~(\ref{eq:rl},\ref{eq:rlt}), still present in Eq.~\eqref{eq:d6sig}, evaluating the T and TT responses only. In the dip region the dominant contribution comes mainly from the $\Delta$ excitation, which is mainly transverse. We have anyway checked numerically that the longitudinal response defined in Eq.~\eqref{eq:rl}  is negligible, amounting to roughly 1\% of the transverse response.

Noteworthy, in the analyzed data the detected outgoing proton lays in the scattering plane, so that $\phi_1'=0,\pi$: $R_{TT}$ component is not affected by the azimuthal angle of the detected particle.

We also performed the isospin separation of the final state into the $pp$ and $pn$ channels. The $pp$ channel arises only by the purely $\Delta$ diagrams, and for these diagrams the ratio between the $pp$ and $pn$ contributions is 1/4, due to the isospin algebra and to the identity of the two protons in the final state.

In Fig.~\ref{fig:NIKHEF1} we show our results for the 2p2h contribution to the $^{12}C(e,e'p)$ cross section, displayed as a function of the missing energy $E_m$ and compared with NIKHEF data from Ref.~\cite{Ryckebusch:1994pi} at electron energy $\varepsilon$=478 MeV, energy transfer $\omega$=263 MeV, momentum transfer $q$=303 MeV/c and proton scattering angle $\theta_p$=38$^\circ$ (first row) and $\theta_p$=113$^\circ$ (second row). 
 In the left panels the separate contributions of the isospin components, corresponding to the emission of two protons ($pp$) and a proton and a neutron ($pn$), are displayed, showing that the contribution of the $pp$ channel is negligible. In the right panel the same cross section is separated into its T and TT components, showing that in this kinematical situation the latter is negative and much smaller than the former. 
The global agreement of the theoretical predictions with the experimental data is very good in the region below the pion production threshold (to be conservative, in the figures we indicate its minimum value, the pion mass).

We note that our results are in qualitative agreement with the calculation of Ref.~\cite{Ryckebusch:1994pi}, performed in a non-relativistic Hartree-Fock finite nucleus model and including final state interactions. The main differences emerge in the position of the 2p2h peak and in the contribution of the $pp$ final state, which in our case is about a factor of 2 smaller.

It is also worth observing that the agreement of our results with those of Ref.~\cite{Ryckebusch:1994pi} indicates that, for these kinematics, relativistic effects are small. A systematical study of relativistic effects in the MEC formalism was performed for the inclusive process in Ref.\cite{DePace:2003spn}, underlining the importance of a relativistic treatment especially for the Delta current and propagator. 

It may be surprising that a model based on the RFG can provide a good description of the 2p2h contribution to the $(e,e'p)$ cross section, whereas it fails in the QE region due to its unrealistic spectral function. 
However, it should be noted that the present approach is based on a pion-correlated Fermi gas, unlike the pure RFG.

In Fig.~\ref{fig:NIKHEF2} some more details of the calculation are illustrated, taking as a reference the $\theta_p$=38$^\circ$ data. In the left panel the separation between direct and exchange contributions shows that the former dominates the response, while the latter is responsible for a slight decrease of the 2p2h peak. 
In the right panel the pionic, $\Delta$ and $\pi-\Delta$ interference are shown: we observe that all contributions are positive, the $\Delta$ alone represents roughly half of the total response, and the purely pionic and $\pi-\Delta$ interference terms are almost equal.

In Fig.~\ref{fig:NIKHEF3} we compare our results with NIKHEF data~\cite{Kester:1995wkm} at another kinematics, for six different values of the proton angle, showing also the separated $pp$ and $pn$ contributions. At lower angles our curves underestimate the data, which are, at these kinematics, more contaminated by quasi-elastic scattering, while at larger angles the comparison is quite good below the pion threshold.

\begin{figure*}[ht!]
    \centering
   \includegraphics[scale=0.35]{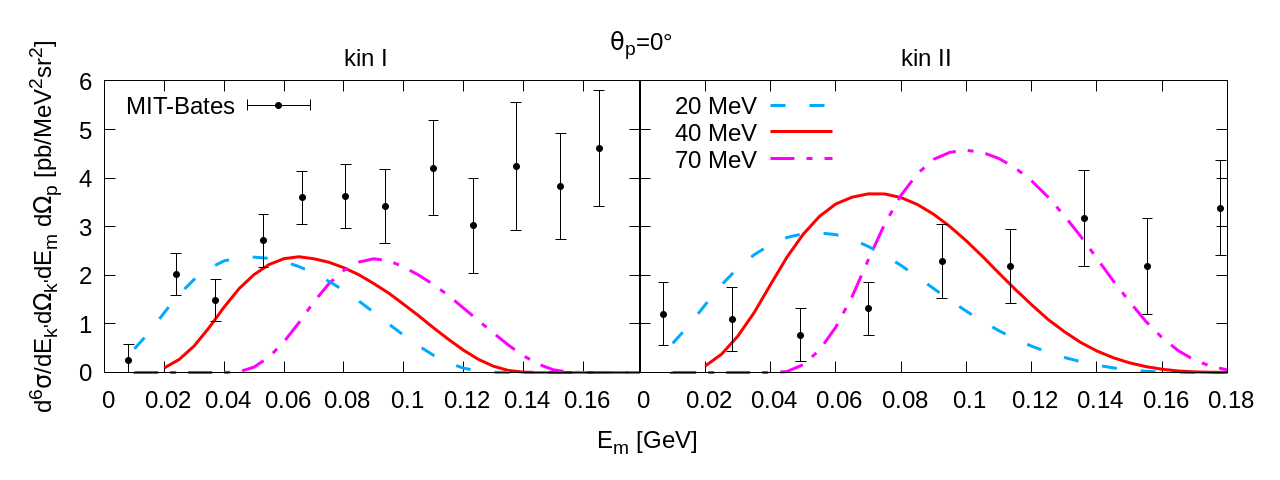 }
    \caption{The 2p2h $^{12}C(e,e'p)$ cross section is displayed versus the missing energy in the parallel kinematic setting ($\theta_p$=0°) for the two kinematics, ``kin I'' ($\varepsilon$=460 MeV, $\omega$=275 MeV, $q$=401 MeV/c) and ``kin II'' $(\varepsilon$=648 MeV,  $\omega$=382 MeV, $q$=473 MeV/c). Data are taken from Ref. ~\cite{Baghaei:1989cc}.
    Different energy shift values have been adopted in the two panels: blue, red and violet curves are obtained using $E_s^{\rm 2p2h}=20,\,40,\,70$ MeV respectively.}
    \label{fig:BATES}
\end{figure*}

In Fig.~\ref{fig:BATES} the 2p2h contribution to the $^{12}C(e,e'p)$ cross section calculated in our model is compared with Bates data~\cite{Baghaei:1989cc} at two different kinematics for proton angle $\theta_p=0$ (the so-called parallel kinematics). In this configuration $W_{xx}^{N}=W_{yy}^{N}$, hence the TT response vanishes and only the transverse response contributes to the cross section. The relative strength of $pn$ and $pp$ pairs, albeit not shown here, is similar to that of Fig.~\ref{fig:NIKHEF1}. 

One can observe that the agreement with data is worse than in the kinematics of Fig.~\ref{fig:NIKHEF1}, especially for the `kin II' data, which are overestimated. A possible interpretation is related to the higher value of $Q^2$: in these conditions nuclear effects not included in our model, such as correlations, could explain the disagreement. It should be noted that, as far as we know, no other theoretical calculation is available at these kinematics to be compared with our result.

\begin{figure*}[ht!]
    \centering
   \includegraphics[scale=0.35]{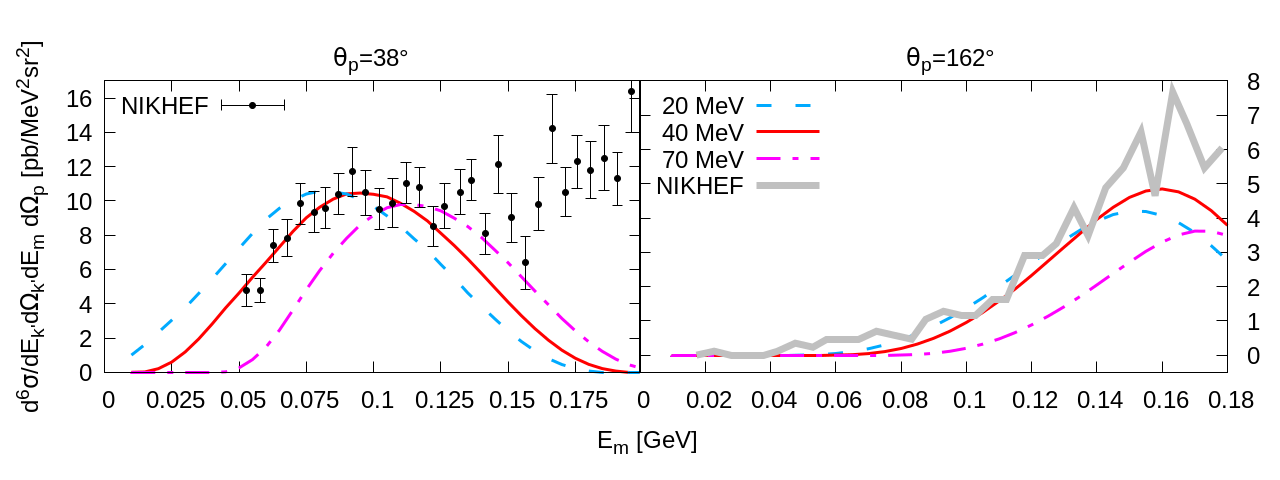 }
    \caption{The same study of the $E_s^{2p2h}$ developed in Fig. \ref{fig:BATES} is performed, but at the kinematics showed respectively in Fig. \ref{fig:NIKHEF1} and Fig. \ref{fig:NIKHEF3}}
    \label{fig:fig6}
\end{figure*}
In this figure and in Fig. \ref{fig:fig6} we also investigate the impact of the energy shift on the theoretical predictions, comparing three values: $E_s^{\rm 2p2h}$=20, 40 and 70 MeV. 
The results of increasing this parameter are a shift towards higher $E_m$ values and a variation in the strength, depending on the analyzed kinematics. The effect is particularly appreciable in the parallel kinematic, where the response is accumulated and localized at lower missing energies. 
In Fig. \ref{fig:fig6} the differences between the curves obtained with the three $E_s^{\rm 2p2h}$ values are smaller: however, in both cases the better agreement seems to be reached with a shift ranging between 20 and 40 MeV. 
It should be noted that $E_s^{\rm 2p2h}$ is, in practice, the only free parameter of the RFG model ($k_F$ being mainly related to the nuclear density). Although a value of 40 MeV in $^{12}$C is quite natural, being twice the shift employed in the quasi-elastic region \cite{Maieron:2001it}, other choices are possible: a shift of 20 MeV has been employed in Ref.~\cite{Megias:2014qva}, whereas 70 MeV is the value yielding the physical separation energy~\cite{VanOrden:2019krz}. Treated as a parameter, $E_s^{\rm 2p2h}$ mimics some of the nuclear effects not yet implemented in the model.

\section{Conclusions}
\label{sec:Conclusions}

We have extended the relativistic model developed in Ref.~\cite{DePace:2003spn} for the 2p2h MEC in inclusive electron scattering, $(e,e')$, to the study of semi-inclusive scattering, $(e,e'p)$. The calculation employs two-body currents involving the exchange of a pion and the excitation of an intermediate $\Delta$ resonance. 
This is the first microscopic calculation of these contributions performed in a relativistic framework.

The model provides a quite satisfactory description of $(e,e'p)$ data 
on $^{12}$C within the kinematic range spanning between the two-nucleon emission and pion production thresholds. A non-trivial and important outcome of the present study is that a description of the 2p2h channel based on the relativistic Fermi gas model is capable of reproducing not only inclusive but also semi-inclusive data. This is at variance with the quasi-elastic region, where the pure RFG is unable to provide a realistic spectral function, resulting inadequate to describe semi-inclusive processes.

The results of the present study give us also confidence in the model's potential applicability to describe more exclusive, flux-integrated neutrino cross sections, such as the one with two-nucleons in the final state, among the most appealing and interesting measurements expected in the near future.

 As Ref.~\cite{DePace:2003spn} represented the reference article for a model subsequently applied to inclusive neutrino cross sections~\cite{RuizSimo:2016rtu,Megias:2016fjk}, the present work aims to play the same role for semi-inclusive neutrino cross sections. Calculations in this direction are in progress.

 The datasets used in this work are unfortunately limited and in some cases affected by large experimental uncertainties. 
However, a reliable modeling of the dip region in terms of the hadronic kinematical variables is mandatory to obtain an adequate description of the spectrum, crucial especially for neutrino scattering. More electron scattering data able to focus on this region could be very useful to explore and test the validity of our model.

\begin{acknowledgements}
This work was partially supported by the Istituto Nazionale di Fisica Nucleare under the National Project "NUCSYS" and by the University of Turin local research funds BARM-RILO-22 and OLIP-DOTTOR-22-03-F. 
The authors acknowledge support from ``Espace de
Structure et de r\'eactions Nucl\'eaire Th\'eorique'' (ESNT,
\url{http://esnt.cea.fr} ) at CEA-Saclay, where this work was partially carried out.
Valerio Belocchi also acknowledges the one-month hospitality of LPNHE Jussieu.
The authors thank Jos\'e Enrique Amaro and Paloma Casal\'e for useful discussions. 
\end{acknowledgements}

\appendix
\section*{Appendix}
In this Appendix we give details on the structure of the two-body currents defined in 
Eqs.~\eqref{eq:pif}, \eqref{eq:seag}, \eqref{eq:deltaF} and \eqref{eq:deltaB}. 
\label{appendix MEC}

\subsection{Pionic currents}
\label{appendix pion}
The pure pionic part of the Electro-Magnetic (EM) MEC is characterized by the exchange of a charged $\pi$ between the two nucleon currents. This appears explicitly in the isospin operator associated to $j^\mu$:
\begin{equation}
    j^\mu_\pi=I_{V_3}J^\mu_\pi \qquad \qquad j^\mu_{\rm sea}=I_{V_3}J^\mu_{\rm sea}
\end{equation}
\begin{equation}
    I_{V}= i \bm \tau^{(1)} \times \bm \tau^{(2)}
\end{equation}
\begin{equation}
    I_{V_3}=\frac{1}{2}\big(\tau_-^{(1)}\tau_+^{(2)}-\tau_+^{(1)}\tau_-^{(2)} \big)
\end{equation}

where $\tau_i$ are the familiar Pauli matrices\footnote{$$\tau_-=\tau_1-i\tau_2=\begin{pmatrix}0 &0 \\ 2 &0\end{pmatrix} \qquad \tau_+=\tau_1+i\tau_2=\begin{pmatrix}0 &2 \\ 0 &0\end{pmatrix}$$}, and the superscripts $(i)$ indicate that the corresponding operator acts on the $i$ particle of the pair~\footnote{This is particularly relevant in the exchange terms computations. In that case particles are inverted, while the operators remain the same, yielding to, for example: 
$$\tau_3^{(1)} |p_2 p_1 \rangle = \tau_3 | p_2\rangle  \otimes |p_1\rangle $$}. 
The $\pi NN$ coupling is of pseudo-vector type, and so the interaction vertex, corresponding to Feynman diagram shown in Fig. \ref{fig:piNNvertex} with ingoing $\pi$, is:

\begin{figure}[htp]
    \includegraphics[scale=0.25]{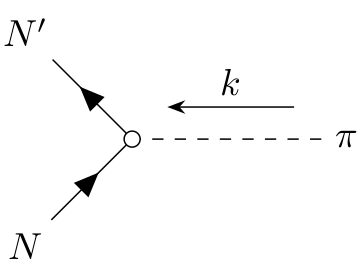}  $\qquad \qquad \qquad $ \includegraphics[scale=0.25]{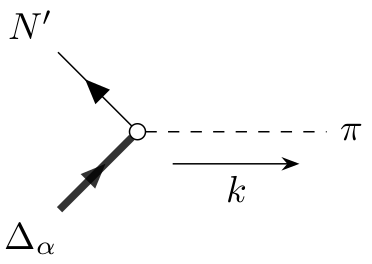}
    \caption{$\pi NN$ coupling with an ingoing pion and $\Delta \rightarrow \pi N$ transition.}
    \label{fig:piNNvertex}
\end{figure}
\begin{equation}
\pi NN\text{ vertex}: \qquad -i\frac{g_A}{2f_\pi}F_{\pi NN}(k^2)\slashed k \gamma_5 \dot{\bm \tau}^\dagger
\end{equation}
$$\dot{\bm \tau} \equiv \Big(\frac{\tau_-}{\sqrt{2}},\frac{\tau_+}{\sqrt{2}},\tau_3\Big) \qquad \text{ coupled with normalized } \bm \pi \equiv (\pi^+,\pi^-,\pi^0)$$
with $g_A=1.26$ the axial constant and $f_\pi=93$ MeV the $\pi$ decay constant.  
Thanks to the Goldberger-Treiman relation it is possible to connect these two quantities with the standard $\pi NN$ coupling $f_{\pi NN}=\sqrt{4\pi\, 0.08}$
\begin{equation}
    \frac{g_A}{2f_\pi}=\frac{f_{\pi NN}}{m_\pi}
\end{equation}
with $m_\pi=139.5$ MeV the $\pi$ mass.

\begin{figure}[htp]
    \includegraphics[scale=0.25]{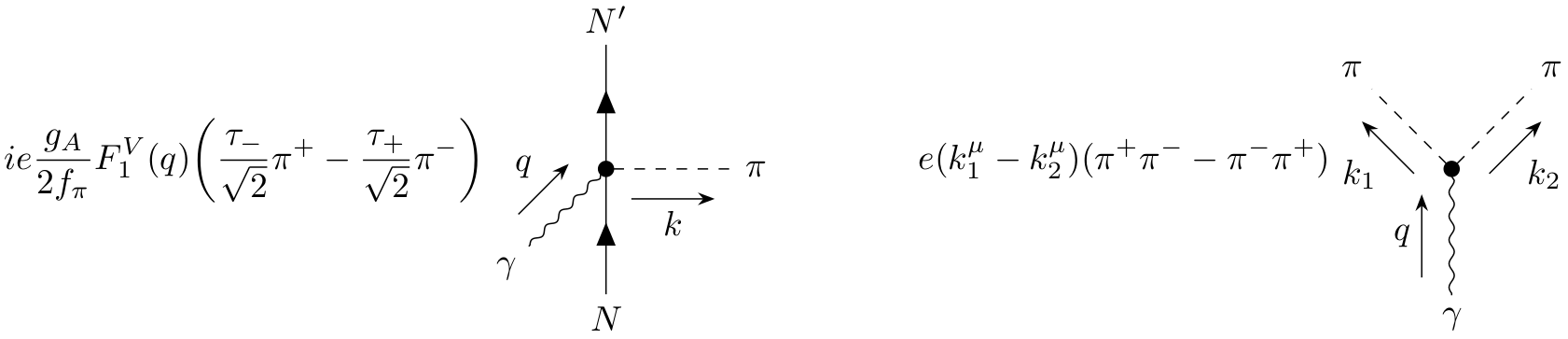}
    \caption{Contact and $\gamma \pi \pi$ EM interaction vertices.}
    \label{fig:piEM}
\end{figure}

The 'contact' EM $\gamma \pi NN$ and the $\gamma \pi \pi$ interactions are shown in Fig. \ref{fig:piEM}, and involve
$F_1^V(q)=F_1^p(q)-F_1^n(q)$ and $e$ the electric charge.\\
The pion appears only as a virtual state for the $2p2h$ process, described through the $\pi$ propagator.
\begin{equation}
    \Delta_\pi(k^2)=\frac{1}{k^2-m_\pi^2}
\end{equation}
To account for the $\pi$ off-shellness in the $\pi NN$ vertex the $F_{\pi NN}$ hadronic form factor is inserted
\begin{equation}
    F_{\pi NN}(k^2)=\frac{\Lambda_\pi^2-m_\pi^2 }{\Lambda_\pi^2-k^2}
\end{equation} 
with $\Lambda_\pi=1300$ MeV.

The $*$ in the pion-in-flight current Eq.~\eqref{eq:pif} stands for the gauge preserving procedure applied, showed in \cite{Gross:1987bu,Dekker:1994yc}, that explicitly is the substitution
\begin{equation}
\big(\Delta_\pi(k_1)\Delta_\pi(k_2)\big)^*=\Delta_\pi(k_1)\Delta_\pi(k_2)+\frac{1}{k_2^2-\Lambda_\pi^2}\Delta_\pi(k_1)+\frac{1}{k_1^2-\Lambda_\pi^2}\Delta_\pi(k_2)
\end{equation}

\subsection{Resonance currents}
\label{appendix delta}

The $\Delta$-MEC include those diagrams in which a nucleon state is excited into a $\Delta$. The transition can happen due to pion or photon interaction. The $\pi N \Delta$ vertex, with corresponding Feynman diagram in Fig. \ref{fig:piNNvertex} representing the $\Delta \rightarrow \pi N$ transition, is:
\begin{equation}
    \pi N \Delta \text{ vertex :} \qquad  i\frac{f^*}{m_\pi}F_{\pi N \Delta}(k^2)k^\alpha \sqrt{\frac{3}{2}}\dot{\bm T}
\end{equation}
$$  \dot{\bm T}:= \Big(\frac{T_-}{\sqrt{2}},\frac{T_+}{\sqrt{2}},T_3\Big) \qquad \text{ coupled with normalized } \bm \pi \equiv (\pi^+,\pi^-,\pi^0)$$
where $T$ is the isospin transition $3/2 \rightarrow 1/2$ operator, which requires the factor $\sqrt{3/2}$ due to its definition\footnote{$$T_1=\frac{1}{\sqrt{6}}\begin{pmatrix}-\sqrt{3} &0 &1 &0 \\ 0&-1 &0 &\sqrt{3} \end{pmatrix} \qquad T_2=-\frac{i}{\sqrt{6}}\begin{pmatrix}\sqrt{3} &0 &1 &0 \\ 0 &1 &0 &\sqrt{3}
\end{pmatrix} \qquad T_3=\frac{2}{\sqrt{3}}\begin{pmatrix}0 &1 &0 &0 \\ 0 &0 &1 &0 \end{pmatrix}$$
$$T^\pm=T_1\pm iT_2$$}. \\$f^*=2.13\times f_{\pi NN}=2.14$ is the coupling constant. The $\Delta$ appears in the MEC just as virtual intermediate state, and $F_{\pi N \Delta}$ is the hadronic form factor that accounts for the off-shell resonance
\begin{equation}
    F_{\pi N\Delta}(k)=\frac{\Lambda_{\pi N\Delta}^2}{\Lambda_{\pi N \Delta}^2-k^2}
\end{equation}
with $\Lambda_\Delta=1150$ MeV.

\begin{figure}[htp]
    \includegraphics[scale=0.25]{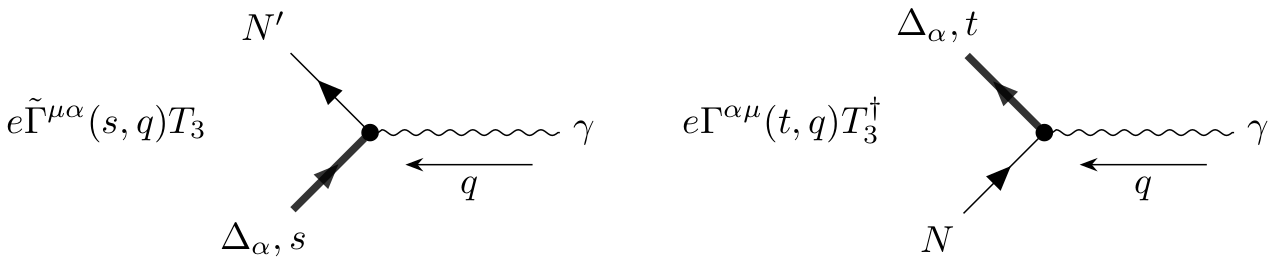}
    \caption{EM $\Delta \rightarrow N$ and $N\rightarrow\Delta$ transition vertices.}
    \label{fig:DeltaEM}
\end{figure}
The $\gamma N \Delta$ interaction vertex is described using a set of form factors which are expansions in the ratio momentum over nucleon mass, inside the definition of $\Gamma^{\alpha \mu}$, that is purely vectorial, as shown in Fig. \ref{fig:DeltaEM}.
In the EM case 
\begin{equation}
    \Gamma^{\alpha \mu}(p,q)=\Big[\frac{C_{3V}(q^2)}{M}(g^{\alpha\mu}\slashed q -q^\alpha \gamma^\mu)+\frac{C_{4V}(q^2)}{M^2}(g^{\alpha \mu}q\cdot p_\Delta-q^\alpha p_\Delta^\mu)+\frac{C_{5V}(q^2)}{M^2}(g^{\alpha \mu}q \cdot p-q^\alpha p^\mu)\Big] \gamma_5
\end{equation}
and
\begin{equation}
   \widetilde{\Gamma}^{\mu \alpha}(p,q)=\gamma^0\Gamma^{\alpha \mu \dagger}(p,-q)\gamma^0 \,,
\end{equation} 
where $M=0.939$ GeV is the nucleon mass and $C_{iV}(q^2)$ are taken from \cite{Hernandez:2007qq}.
The dominant form factor is related to the $C_{3V}$ term.\\

$\Delta$-MEC involve only the virtual resonance, whose propagation is described by the Rarita-Schwinger propagator
\begin{equation}
G_{\alpha \beta}(p)=\frac{\mathcal{P_{\alpha \beta}}(p)}{p^2-M_\Delta^2+iM_\Delta\Gamma_\Delta(p)}
\,,\label{eq:prop}
\end{equation}
where $\mathcal{P_{\alpha \beta}}$ is the projector over the physical states \footnote{Note that $$\overline{\mathcal{P}_{ \beta \alpha}}(p) := \gamma^0 \mathcal{P_{\alpha \beta}}^\dagger(p)\gamma^0=\mathcal{P_{\beta \alpha}}(p)\,.$$}
\begin{align}
\sum_{spin}&u_\alpha(p)\overline{u}_\beta(p)=\mathcal{P_{\alpha \beta}}(p) = -(\slashed p+M_\Delta)\Big[g_{\alpha\beta}-\frac{1}{3}\gamma_\alpha \gamma_\beta-\frac{2}{3}\frac{p_\alpha p_\beta}{M_\Delta}+\frac{p_\alpha\gamma_\beta-p_\beta\gamma_\alpha}{3M_\Delta}\Big] \,.
\end{align}
The $\Delta$ propagator (\ref{eq:prop}) takes into account the vacuum width decay $\Gamma_\Delta$ for the pion production process, following \cite{Dekker:1994yc},  but in the computation only the real part of the propagator is included. This procedure, also adopted in Ref.~\cite{DePace:2003spn}, is an effective way to avoid double counting with pion production process.

The free $\Delta$ decay width is
\begin{equation}
\Gamma_\Delta(p)=\frac{(4 f_{\pi N \Delta})^2}{12\pi m_\pi^2} \frac{|\mathbf{k}|^3}{\sqrt{p^2}} (M + E_k) F(k_{\rm rel}^2)
\end{equation}
where $(4 f_{\pi N \Delta})^2/(4\pi)=0.38$, $p^2$ is the $\Delta$ invariant mass, $\mathbf{k}$ is the produced pion or nucleon three-momentum in the $\Delta$-at-rest frame, such that
\begin{equation}
\mathbf{k}^2=\frac{1}{4p^2}[p^2-(M+m_\pi)^2][p^2-(M-m_\pi)^2]\,
\end{equation} 
and $E_k=\sqrt{M^2 + \mathbf{k}^2}$ is the associated nucleon energy.
In order to better reproduce experimental data, the additional factor
\begin{equation}
F(k_{\rm rel}^2)=\left(\frac{\Lambda_R^2}{\Lambda_R^2-k_{\rm rel}^2}\right)
\end{equation}
is considered, with $k_{\rm rel}^2=(E_k - \sqrt{m_\pi^2 + \mathbf{k}^2})^2-4\mathbf{k}^2$ the relative $\pi-N$ four-momentum and $\Lambda_R^2=0.95\, M^2=0.915^2$ GeV$^2$.~\footnote{In \cite{Dekker:1994yc}, in the denominator of Eq.~(A5), $\Lambda^2$ appears instead of $\Lambda_R^2$.}

Isospin $\Delta$-MEC operators and explicit currents are reported here~\footnote{Note that in Eq.(63) of Ref.~\cite{Rocco:2018mwt} the factor $3/2$ in the $\Delta$ current should read $\sqrt{3/2}$.}.
Note that in this case the exchange $1\leftrightarrow 2$ of Eqs.~\eqref{eq:deltaF}, \eqref{eq:deltaB} acts on $I_\Delta$ too, changing the $I_{V_3}$ part in sign:
\begin{equation}
    I_{\Delta_{\rm F}}=-2\tau_3^{(1)}-I_{V_3} \qquad I_{\Delta_{\rm B}}=-2\tau_3^{(1)}+I_{V_3}\,.
\end{equation}

\begin{equation}
    J^\mu_{\Delta_1} = \frac{f^*f_{\pi NN}}{\sqrt{6}m_\pi^2} F_{\pi N\Delta}(k_1)F_{\pi NN}(k_1) \bar{u}(p_1)\gamma_5 \slashed{k_1} u(h_1)\Delta_\pi(k_1) k_1^\alpha  \bar{u}(p_2)\Big[G_{\alpha \beta}(t_2)\Gamma^{\beta \mu}(h_2,q)+\widetilde{\Gamma}^{\mu \beta}(p_2,q)G_{\beta \alpha }(s_2)\Big] u(h_2)
\end{equation}
\begin{equation}    
J^\mu_{\Delta_2} = J^\mu_{\Delta_1}(1 \leftrightarrow 2)
\end{equation}
\begin{eqnarray}
J^\mu_{\Delta_3} = \frac{f^*f_{\pi NN}}{\sqrt{6}m_\pi^2} F_{\pi N\Delta}(k_1)F_{\pi NN}(k_1) &\bar{u}(p_1)\gamma_5 \slashed{k_1} u(h_1)\Delta_\pi(k_1)k_1^\alpha  \bar{u}(p_2)\Big[G_{\alpha \beta}(t_2)\Gamma^{\beta \mu}(h_2,q) -\widetilde{\Gamma}^{\mu \beta}(p_2,q)G_{\beta \alpha }(s_2)\Big] u(h_2) \nonumber \\
\qquad - (1 \leftrightarrow 2)&
\end{eqnarray}

\bibliography{bib_mec}
\end{document}